\documentclass[12pt,preprint]{aastex}
\usepackage{emulateapj5}
\usepackage{graphicx}

\shorttitle{$\pi$ in the sky}
\shortauthors{Holz \& Wheeler}

\begin{document}

\title{Retro-MACHOs: $\pi$ in the sky?}
\author{Daniel E. Holz}
\affil{Institute for Theoretical Physics, University of
California, Santa Barbara, CA 93106}
\and
\author{John A. Wheeler}
\affil{Department of Physics, Princeton University, Princeton, NJ 08544}

\begin{abstract}
Shine a flashlight on a black hole, and one is greeted with
the return of a series of concentric rings of light. For a
point source of light, and for perfect alignment of the
lens, source, and observer, the rings are of infinite
brightness (in the limit of geometric optics). In this
manner, distant black holes can be revealed through their
reflection of light from the Sun. Such retro-MACHO events
involve photons leaving the Sun, making a $\pi$ rotation
about the black hole, and then returning to be detected at
the Earth. Our calculations show that, although the light
return is quite small, it may nonetheless be detectable for
stellar-mass black holes at the edge of our solar
system. For example, all (unobscured) black holes of mass
$M$ or greater will be observable to a limiting magnitude
$\bar{m}$, at a distance given by:
$0.02\,\mbox{pc}\times\sqrt[3]{10^{(\bar{m}-30)/2.5}\,(M/10\,M_\sun)^{2}}$.
Discovery of a Retro-MACHO offers a way to {\em directly}\/ image the
presence of a black hole, and would be a stunning
confirmation of strong-field general relativity.
\end{abstract}


\section{Introduction}

In the discovery of MACHOs light has shown its power to
reveal dark compact objects. In these events the photon from
a distant source suffers a very small angular deflection,
small enough to make gravitational lensing the relevant
mechanism.  The bending power of a black hole is not
limited, however, to small angles but reaches to $\pi$ and
odd multiples of $\pi$. Illuminated by a powerful point
source of light, the black hole therefore will shine back
with a series of concentric rings (we call this
retrolensing). A waterdrop, too, likewise illuminated,
shines back, but for a different reason: the internal
reflections experienced by the photons. That returned light
shows to the air traveler flying over a fogbank as a
``glory'': a rainbow-like halo surrounding the shadow of the
plane on the cloud. Each ray of light that contributes to
this sensation has suffered its $\pi$ deflection in a
different waterdrop, therefore the impression of colored
rings that the glory makes is impression only, built up in
the last analysis out of multitudes of tiny dots of
illumination. No one would be so rash as to expect a
detectable backscatter from a single water droplet. It is
precisely the search for such backscatter from a single
black hole (the putative retro-MACHO) that is,
however, the topic of this paper.

How search for retro-MACHOs, for ``$\pi$ in the sky'',
especially when dogged by the negative implications of the
familiar phrase? To tell observers to go and look everywhere
is to them as dismaying as to be told to look
nowhere. Fortunately the successful search for MACHOs
provides a helpful model, and the records from that search
an immediate place of reference. Recent microlensing events appear
to directly confirm that a population of stellar-mass black
holes exists in our galaxy~\citep{macho2,macho3,agol2}.  The Sun, in
turn, furnishes a powerful nearby source with whose help one
can hope to search a nearby region of the galaxy for black
holes not otherwise revealed, taking advantage of times when
the Earth interposes itself to spare the registering device
from the direct glare of the light source. Although the
observation of a retro-MACHO is unlikely, it
nonetheless affords one of the few ways to directly image
nearby black holes, and if ever observed, would make for an
impressive confirmation of Einstein's theory.

\section{What is a retro-MACHO?}

Consider a point source of light, an observer, and a black
hole lens, with all three co-linear and the observer located
{\em in between} the source and black hole. The observer,
looking at the image of the source in the black hole, will
see (resulting from symmetry) an (infinite) series of concentric
rings, corresponding to bending angles of successively
higher odd multiples of $\pi$. The outermost ring would be
from photons suffering a $\pi$ deflection, the next inner
ring corresponding to a $3\pi$ deflection, and so on. For a
point source, and in the limit of geometric optics, the
amplification of the rings in infinite. For realistic
scenarios with extended sources, however, the amplification
is finite.  For example, consider an extended spherically
symmetric source emitting uniformly across its surface. The
image of the source at the observer is an annulus, where the
outer and inner radii correspond to the appropriate impact
parameters for photons coming from the top and bottom parts
of the source, as shown in Fig.~\ref{perfect_align}. Since
lensing conserves surface brightness, the brightness of the
ring image relative to that of the source is given by the
ratio of the areas of the images, as seen at the observer.

The fraction of the sky covered by the ring image is given by:
\begin{equation}
A_I={\pi{b_o}^2-\pi{b_i}^2\over4\pi {D_{OL}}^2},
\end{equation}
where $b_i$ and $b_o$ are the impact parameters corresponding to the
inner and outer radii of the ring, respectively. Note that the impact
parameter is not the same as the periastron (closest approach)
distance:
\begin{equation}
b={r\over\sqrt{1-{2M/r}}},
\end{equation}
where $b$ is the impact parameter, $r$ is the periastron distance, and
$M$ is the mass of the black hole. Throughout we measure all
quantities in geometric units, with $G=c=1$ (e.g.,
$1\,M_\sun=1.5\,\mbox{km}$).

This is to be compared with the sky coverage of the source,
as seen directly by the observer:
\begin{equation}
A_S={\pi R_S\/^2\over4\pi D_{OS}\/^2}.
\end{equation}
The total amplification, $\mu$, of the image is given by the ratio
of their sky coverages, yielding
\begin{eqnarray}
\mu&=&{b_o\/^2-b_i\/^2\over {R_S}^2}\left({D_{OS}\over
D_{OL}}\right)^2.
\label{eq:mu_1}
\end{eqnarray}

To calculate the brightness of the image we need to know the
inner and outer radii of the ring, represented by rays from
points $A$ and $B$ in Fig.~1, respectively. Rays from $A$
suffer a deflection angle of $\pi-\alpha$, while for those
from $B$ the deflection is $\pi+\alpha$, where
$\alpha={R_S/D_{LS}}$.  In everything that
follows, we restrict our attention to the primary
(outermost, brightest) images, although a whole succession
of images is present (corresponding to higher odd multiples
of $\pi$). For the sake of simplicity, we confine ourselves
to the case of non-rotating (Schwarzschild) black holes. The
calculations for rotating (Kerr) holes are significantly
more complex~\citep{rb94}, and will be deferred to later work. To
calculate the impact parameters that correspond to the
necessary deflection angles, we take advantage of an
approximate expression in \citet{luminet} (see also
\citet{chandra}) for the case of
large deflection:
\begin{equation}
b=b_c+b_d\,e^{-\Theta},
\label{chandra_approx}
\end{equation}
where $b_c=3\sqrt3\,M$,
$b_d=648\sqrt3 e^{-\pi}\,M (\sqrt3-1)^2/(\sqrt3+1)^2
\approx3.4823\,M$,
and $\Theta$ is the deflection angle.
Chandrasekhar also provides exact analytic forms for the solution
(in terms of Elliptic functions). We find that the
approximate expressions are within $10\%$ of the exact
values for bend angles satisfying $\Theta\ge\pi/4$, and are good to
within $1/2\%$ for bend angles with $\Theta\ge\pi$.

Using the results of eq.~\ref{chandra_approx} to calculate
$b_o$ and $b_i$, eq.~\ref{eq:mu_1} can be simplified to
\begin{equation}
\mu\simeq 3.22\,M^2\,{D_{OS}\/^2\over R_S D_{LS}D_{OL}\/^2}.
\label{result1}
\end{equation}
Eq.~\ref{result1} assumes perfect (colinear) alignment of
the source, observer, and lens. The generic case, of course,
will involve misalignment, as shown in Fig.~2. We define the
misalignment angle, $\beta$, to be the angle between the
source--observer line and the observer--lens line, as
measured at the observer (such that $\beta=0$ corresponds to
the case of perfect alignment shown in Fig.~1).  As $\beta$
increases from zero the ring

\hspace*{0cm}\resizebox{8.5cm}{!}{
\includegraphics{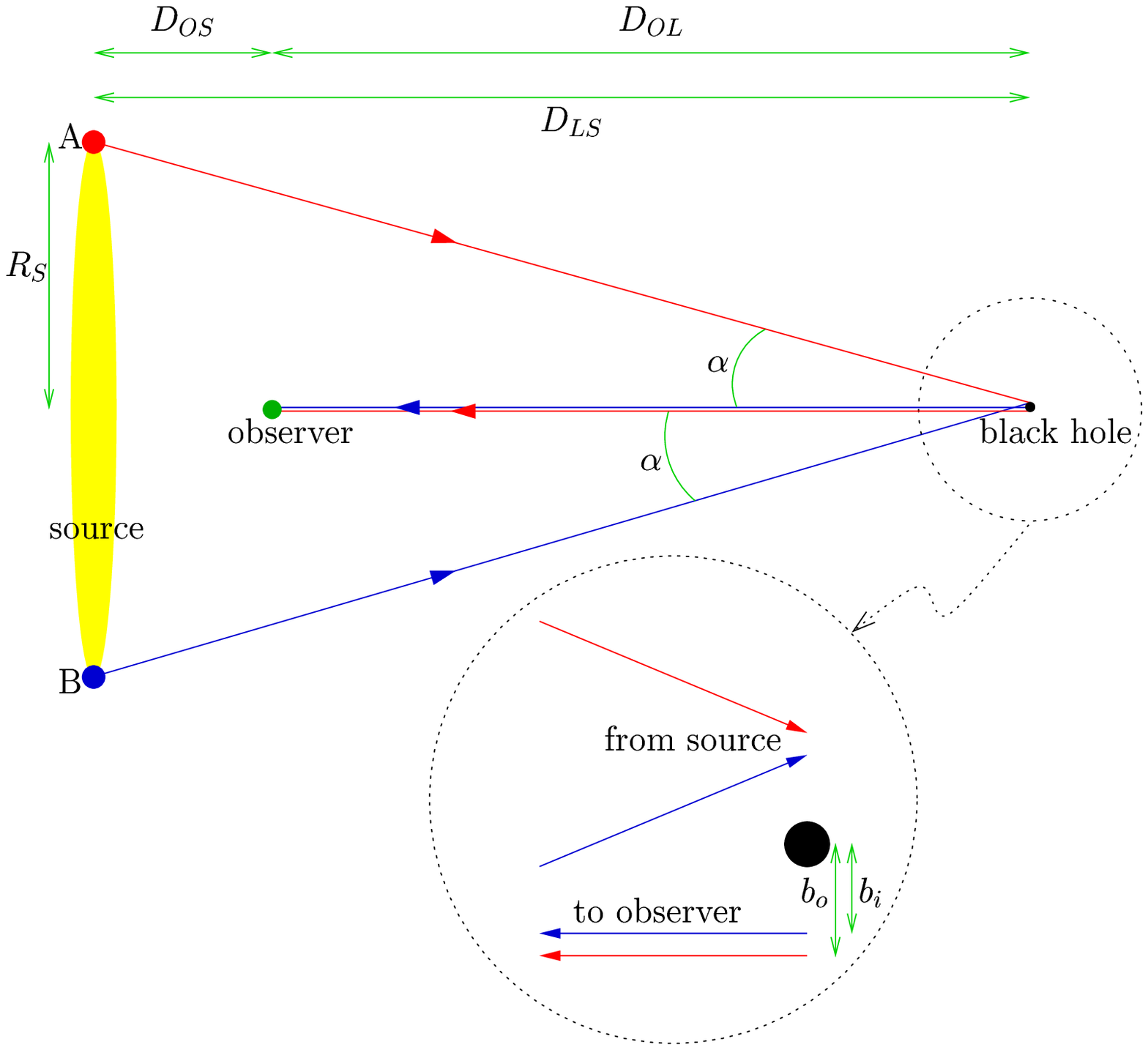}}
\figcaption[perfect_align.eps]{Perfect alignment: the
(extended) source, observer, and lens are colinear. The resulting image
of the source, as lensed by the black hole, is a ring. (The
angles in this figure are greatly exaggerated.)
\label{perfect_align}}
\vspace{0.25cm}
\noindent
splits into two arcs (on either
side of the lens), centered on the lens--observer--source
plane (see the inset of Fig.~2), with the arcs shrinking to
small, point-like images as $\beta$ approaches $\pi/2$. The
tangential extent of these arcs, as seen at the observer, is
determined by the angular extent of the protrusion of the
source out of the plane of Fig.~2: $\Delta\Theta\approx2
\tan^{-1}\left({R_S/D_{OS}\sin\beta}\right)$. The total
area of each image is thus found to be
\begin{eqnarray} A_{\rm
image}&=&\pi(b_o\/^2-b_i\/^2)\times{\Delta\Theta\over2\pi}\\
&=&(b_o\/^2-b_i\/^2)\tan^{-1}\left({R_S\over D_{OS}\sin\beta}\right),
\end{eqnarray} where we have neglected the fact that the
image heights taper down to zero at their edges. Here
$b_o$ and $b_i$ represent the outer and inner apparent
impact parameters of the images, which correspond to bend
angles of $(\pi -\delta)\mp\alpha$ for the
bigger image, and $(\pi+\delta)\mp\alpha$ for the smaller
image. Note that, since surface brightness is conserved, the
bigger image is also the brighter one (and is found further
from the center of the lens, on the side opposite that of
the source). The magnification of each
image is again given by the ratio of the areas of the
source and image, namely:
\begin{equation}
\mu=(b_o\/^2-b_i\/^2)\tan^{-1}\left({R_S\over
D_{OS}\sin\beta}\right){D_{OS}\/^2\over\pi R_S\/^2
D_{OL}\/^2}.
\label{result2}
\end{equation}
In the limit $\beta\rightarrow 0$, and summing up both
images, eq.~\ref{result2} goes over to eq.~\ref{eq:mu_1}, as
expected.  Using eq.~\ref{chandra_approx}, and taking both
$\alpha\ll 1$ and $\beta\ll 1$, the combined magnification of both
images simplifies to
\begin{equation}
\mu\simeq 3.22\,M^2\,{D_{OS}\/^2\over R_S D_{LS}D_{OL}\/^2} 
- 2.05\,\beta\,M^2\,{D_{OS}\/^3\over R_S\/^2 D_{LS}D_{OL}\/^2}.
\label{result3}
\end{equation}

\section{Sources for retro-MACHOs}

Every source of light on the sky has the potential to be
lensed by a retro-MACHO (in the same manner that
\hspace*{0cm}
\resizebox{8.5cm}{!}{ \includegraphics{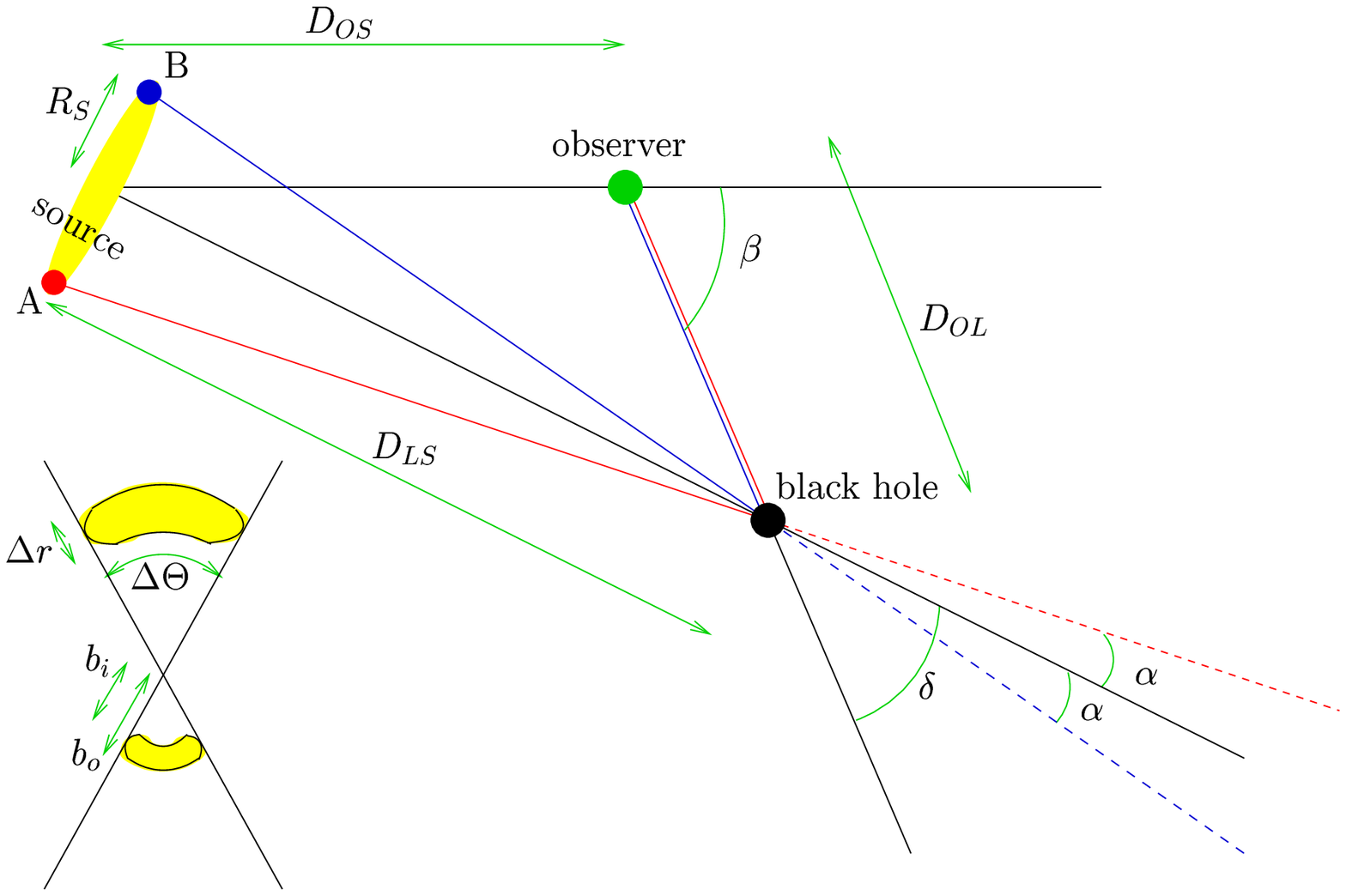}}
\figcaption[perfect_align.eps]{Imperfect alignment: the source,
observer, and lens are not colinear. Pairs of images are
produced, centered on the source--observer--lens plane, on opposite
sides of the lens (see inset).
\label{imperfect_align}}
\vspace{0.25cm}
\noindent
the entire
Universe is imaged near the horizon of each and
every
(unobscured) black hole). However, given the very slight
magnification, only the very brightest of sources, with the
most felicitous of alignments, will be observable.

We envisage the retro-MACHO to sit at one or another
distance and ask for its expected luminosity in order to
compare it with familiar objects in the sky. In each case we
take the primary source of illumination to be the Sun
(output $I= 4\times10^{33}\,\mbox{erg}/\mbox{sec}$, apparent
visual magnitude $m=-26.8^m$).  In Table~1, we show the
observed apparent visual magnitudes of these hypothetical solar
retro-MACHOs. In addition to perfect alignment ($\beta=0$),
it is also useful to consider the case of edge alignment
($\beta=R_S/D_{OS}=R_\sun/1\,\mbox{AU}$), wherein the observer--lens line grazes
the edge of the source. Because of its large angular size
(as seen from Earth), the Sun offers the best opportunities
for close alignment. In Eq.~\ref{result2} we find that the
Sun's $1/2\degr$ angular size 
dominates over smaller values of $\beta$ for other sources.

A number of other potential sources, in addition to the Sun,
are worth mentioning. Consider Sirius, the next
brightest star in the sky. For a perfectly-aligned
retro-MACHO of $1\,M_\sun$, at a distance of
$0.01\,\mbox{pc}$, the retro-image of Sirius would appear at
$34^m$. A misalignment of $1\degr$ would put its brightness
at $51^m$. The reg giant Betelgeuse would have a brightness
of $39^m$ perfectly aligned, and drop to $53^m$ when
$\beta=1\degr$. These are to be compared to the Sun, which
has a retro-image at $31^m$ for perfect alignment. For
$\beta=1\degr$ this drops by only 3 mag, to $34^m$,
showing the advantage of the Sun's large angular size.

As the magnification factors in eqs.~\ref{result1}
and~\ref{result3} scale as $M^2$, one might be tempted to
consider the supermassive black hole at the galactic center
as a retro-MACHO lens. The Sun, at its closest ($5\degr$)
alignment with the galactic center black hole, has a
retro-image brightness of $50^m$. This becomes even fainter
when the considerable extinction to the galactic center
(double the usual value, as the photons travel round trip)
is taken into account.

It is to be noted that all of the expressions found in this
paper become relevant for black holes at cosmological
distances by replacing the physical distances with the
appropriate cosmological luminosity distances.  One can
imagine imaging the Milky Way galaxy in the supermassive black
hole of a nearby galaxy.
\hspace*{0cm}\resizebox{8.5cm}{!}{
\includegraphics{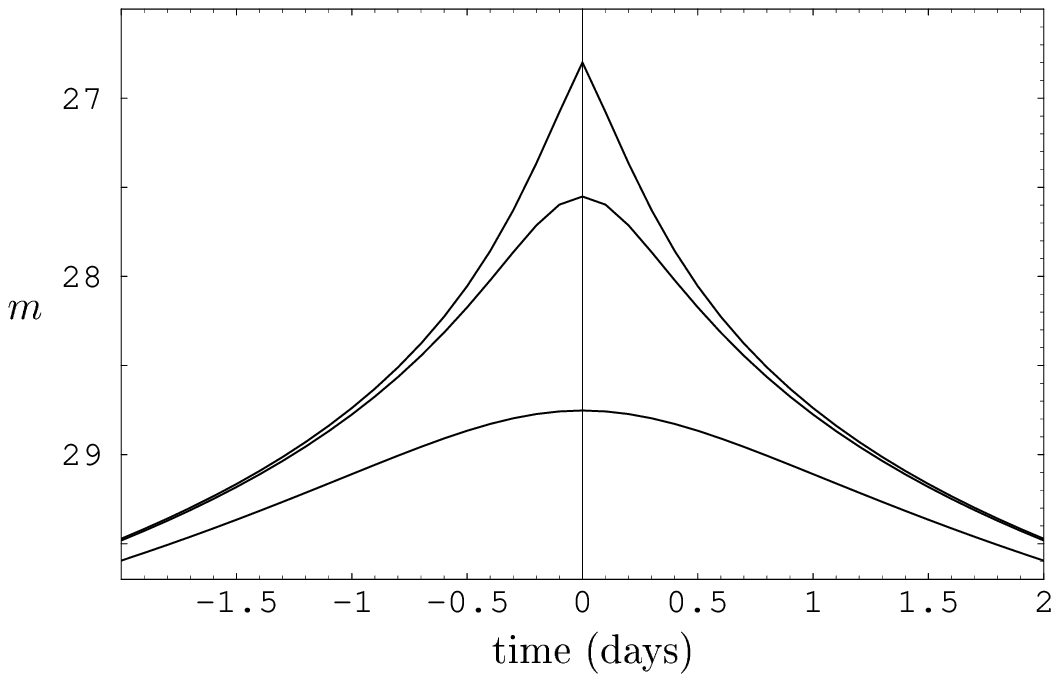}}
\figcaption[lightcurve1.eps]{Solar retro-MACHO lightcurves: The
apparent visual magnitude, $m$, of the Sun, imaged in a
$10\,M_\sun$ black hole at a distance of
$0.01\,\mbox{pc}$. The different curves are for the black
hole at angular displacements from the ecliptic plane of 0, $R_\sun/1\,\mbox{AU}$,
and $1\degr$ respectively (top to bottom).
\label{lightcurves}}
\vspace{0.25cm}
\noindent
However, even if Andromeda had an
unobscured central supermassive black hole, the image
of the
Milky Way would be dimmer than $70^m$. In more
extreme cases, we could imagine a supernovae or quasar
imaged (in any frequency band, e.g. radio) in a distant
supermassive black hole. Even for perfect alignments, such
scenarios produce extremely dim images.

\section{Properties of retro-MACHOs}

A retro-MACHO would appear somewhat like a conventional
MACHO. However, rather than magnifying a pre-existing star,
a retro-MACHO causes a ``star'' to appear out of
nowhere. A number of sample retro-MACHO lightcurves are
shown in Fig.~\ref{lightcurves}. The colors of a retro-MACHO are identical to those
of the source, and this will provide an important way to
confirm the nature of such events. The structure of a
retro-MACHO will be unresolvable (much like the multiple
images in microlensing are unresolvable), as the radius of
the ring is on the order of the Schwarzschild radius of the
lens. Even for very close black holes (say, at the edge of
the solar system), the angular extent of a retro-MACHO image
remains less than a milliarcsec.

As in the MACHO case, the timescale of retro-MACHO
events is determined by the relative motion of the source,
lens, and observer. However, rather than the Einstein angle
of the lens being the relevant angular scale, in the retrolensing
case we are interested in the angular extent of the source
measured at the observer. We define the timescale of an
event to be given by the length of time it takes to go from
perfect alignment to edge alignment:
\begin{equation}
t={D_{OL}\over v} {R_S\over D_{LS}},
\end{equation}
where $v$ is the transverse velocity of the observer with
respect to the source--lens line-of-sight. The Sun's angular
extent is $1/2\degr$, which means that a solar retro-MACHO
lasts about $1/2$ day, assuming the source--lens motion is
stationary as compared to the motion of the Earth about the
Sun.
For lens motion at galactic velocities
($\sim200\,\mbox{km}/\mbox{sec}$), the timescale for the
lens to move an appreciable angle on the sky is considerably
longer (unless the lens is within the solar system). For the
case of distant stars being retro-lensed by distant black
holes, however, the timescales can shorten because of the extremely
small angles needed for edge alignment. The retro-image of
Sirius in a black hole about a parsec away can be expected
to last on the order of hours, with the precise value
depending on the relative projected velocities.

Another point of consideration is the light travel-time of a
given retro-MACHO event. For example, for a solar
retro-MACHO at a distance of a parsec, the event will appear
on the sky corresponding to a position directly opposite to
where the Sun was located $2\,\mbox{pc}/c\approx 6\
\mbox{years}$ ago! For a retro-MACHO to occur within
$1\degr$ of the current antipodal position of the Sun, the
black hole must be within about a day's round-trip light
travel time, putting the black hole lens at the distance of
Pluto.

In addition, solar retro-MACHO events will repeat
annually, as the Earth returns to the appropriate alignment
spot in the ecliptic. This annual repetition may become one
of the most valuable signatures of retro-MACHOs, and can be
used both to identify candidates in sky surveys, and to
confirm the nature of the events.

\section{Observing strategies}

From Table~1, it is apparent that retro-MACHOs are
exceedingly dim, even in the best of cicumstances. The most
likely observed retro-MACHO would be our Sun imaged in a
nearby stellar-mass black hole. Simple theoretical estimates
give a value for the local number density of stellar-mass
black holes of
$N_{\mbox{\footnotesize BH}}\sim8\times10^{-4}/\mbox{pc}^{3}$
\citep{st}. Based on the microlensing event rates towards the galactic
bulge of the MACHO and OGLE teams, we find
$N_{\mbox{\footnotesize
BH}}\sim2\times10^{-4}/\mbox{pc}^3$~\citep{agol,macho1}. Both
these rates include black holes in the mass range
$5\,M_\sun\la M_{\mbox{\footnotesize
BH}}\la15\,M_\sun$. Although it is unlikely for there to be
a stellar-mass black hole within a parsec, it is reasonable
to expect one within $10\,\mbox{pc}$ of the Sun. Such a
black hole, if perfectly aligned with the Sun (i.e. in the
ecliptic plane), would produce a $49^m$ retro-image of the
Sun. Even at a distance of $1\,\mbox{pc}$, the image only
brightens to $41^m$. The earliest population of stars are
thought to have masses in the 300--1,000$\,M_\sun$ range,
and probably collapse to form black holes in a similar mass
range (see, e.g.,~\citet{fhh}). The retro-MACHOs from these
holes would be visible at $32^m$ to a distance of over a
parsec. It is apparent, however, that even in the best of
circumstances only a chance encounter with an unusually
nearby or massive black hole will lead to an observable
retro-MACHO.

Microlensing surveys have produced a number of candidate
black hole lenses. These arise from long-timescale events,
where the parallax effect of the Earth's revolution about
the Sun is strong enough to break some of the degeneracies
in the microlensing fits. For example, the best fit to MACHO
galactic bulge event 96-BLG-5 is a black hole of mass
$6\,M_\sun$ at a distance of $1\,\mbox{kpc}$~(Bennett et
al. 2001; see also Agol et al. 2002). Although this is too
distant to be observed, it is conceivable that a future
microlensing event may provide us with a definite
retrolensing candidate. In these cases we know exactly where
and when to look, and what to expect.

Although the likelihood is small, a close approach of a
stellar-mass black hole would have a potentially
catastrophic impact on our solar system (both through
disruption of the Oort cloud, and through direct effects on
the orbital stability of the planets). Retrolensing offers
perhaps the only {\em direct}\/ observational signature of
such an incoming ``rogue'' black hole.  The Sun can be
considered a searchlight, sweeping out a $1/2\degr$-wide
swath of the sky centered on the ecliptic plane in search for
black holes. For a magnitude limit of $\bar{m}$, this reveals
all black holes of mass $M$ or greater out to a limiting
distance of
$0.02\,\mbox{pc}\times\sqrt[3]{10^{(\bar{m}-30)/2.5}\,(M/10\,M_\sun)^{2}}$.
Through repeated, deep observations of the direction
anti-podal to the Sun, retrolensing can be utilized to
construct a black hole early warning system.\footnote{
It is to be noted that a very powerful ($>\mbox{ Gigawatt}$) laser can be used to
{\em actively}\/ probe for retro-MACHOs. One could scan the entire
sky systematically, and monitor for returning glints of
light (in the same frequency band as that of the laser, and
with time delays a direct measure of distance to the black
holes). This would constitute a true black hole early
warning system.}
Consider
Nemesis, a hypothetical solar companion~\citep{hills}. If we
take Nemesis to be a $1\,M_\sun$ black hole at the outer
edge of the Oort cloud ($D_{OL}=9\times 10^4\,\mbox{AU}$),
and lying near the ecliptic, it would be directly imaged as
a retro-MACHO at $43^m$. However, such a Nemesis {\em
would}\/ be revealed ($m<30^m$) if it approached to a distance of
$D_{OL}\la10^3\,\mbox{AU}$.
It is to be noted that microlensing, although quite
proficient at detecting compact objects in the direction of
the galactic bulge at distances of kiloparsecs, is not
particularly efficient at detecting nearby black
holes. The Einstein angle of a point mass lens is given by
$\theta_E=\sqrt{4MD_{LS}/(D_{OL}D_{OS})}$, which for distant stars
lensed by nearby masses simplifies to
$\theta_E\approx\sqrt{4M/D_{OL}}$. For a black hole at
$D_{OL}=10^3\,\mbox{AU}$, a star would have to be aligned to
within $1.5\,\mbox{arcsec}$ to be microlensed. For a solar
retro-MACHO, on the other hand, edge-alignment is
automatically satisfied for any black hole lying within
$1/4\degr$ of the ecliptic. Retrolensing is thus better than
microlensing for detecting approaching nearby
black holes.

We are entering an age where the entire sky can be
expected to be imaged deeply, and all transient objects
noted~\citep{blandford}. Within any such large, repetitive
sample, a retro-MACHO would look like a MACHO event with
zero baseline flux (i.e. no background star). Upon detecting
such an event, one would try to identify a possible light
source at the diametrically opposite position on the sky
(``through'' the Earth). Does the source have identical
colors to the transient image? If so, retro-MACHO!  For
example, in the case of the Sun, we would look for images
with precisely solar spectra. It may be of interest to
perform a thorough search of extant microlensing survey
databases for events that do not correspond to
amplification of background stars. Presuming a non-detection
of such events, and taking the depth of current surveys to
be $\sim22^m$, we can directly rule out unobscured black
holes with mass greater than $10\,M_\sun$ in the direction
of the galactic bulge ($\beta\ge5\degr$) at distances less
than $150\,\mbox{AU}$. These distance limits improve
dramatically with deeper surveys.

\smallskip
Should a retro-MACHO ever be observed, it would afford an
unprecedented opportunity to study the strong-field aspects
of general relativity. These objects probe very near the
horizon of the holes---in the perfect alignment case, the
image represents photons that have passed
within $ 3.5 m=1.75 r_s$ of the singularity (where $r_s$ is the
Schwarzschild radius), as deep into the strong field regime
as we are ever likely to probe via light. A retro-MACHO
observation would be an impressive confirmation of the
theory of general relativity, and would leave little doubt
as to the existence of Einstein black holes.



\acknowledgments
It is a pleasure to acknowledge discussions with Eric Agol,
Doug Eardley, Warner Miller, Bohdan Paczy\'nski, James
Peebles, Sterl Phinney, Bob Rutledge, and Johndale Solem in
the course of this work. We thank Emily Bennett for
providing invaluable assistance in the preparation of the
manuscript.

\begin{deluxetable}{ccccccc}
\tablecaption{retro-MACHO brightnesses of the Sun\label{tbl-1}}
\tablewidth{0pt}
\tablehead{
\colhead{BH mass} & \colhead{BH distance}   & \colhead{$\beta=0$}
& \colhead{$\beta=R_\sun/1\,\mbox{AU}$}  & \colhead{$\beta=1\degr$}  
& \colhead{$\beta=\pi/4$} & \colhead{$\beta=\pi/2$} \\
\colhead{($M_\sun$)}  & \colhead{(pc)}   & \colhead{(perfect alignment)}
& \colhead{(edge alignment)}  & 
&  & (max misalignment)
}
\startdata
1 & $10^{-2}$ & 31.0 & 32.6 & 34 & 38 & 38\\
1 & $10^{-1}$ & 38.6 & 40.1 & 41 & 45 & 46\\
10 & $10^{-2}$ & 26.1 & 27.6 & 29 & 33 & 33\\
10 & $10^{-1}$ & 33.6 & 35.1 & 36 & 40 & 41\\
10 & 1 & 41.1 & 42.6 & 44 & 48 & 48
\enddata


\end{deluxetable}

\end{document}